\documentclass[fleqn,useAMS,usenatbib,twocolumn]{aastex701}
\usepackage{amsmath}
\usepackage{graphicx,color,hyperref}
\usepackage{ulem}
\AtBeginDocument{
  \setlength{\parskip}{0pt}
  \setlength{\belowdisplayskip}{4pt}
  \setlength{\belowdisplayshortskip}{4pt}
}

\renewcommand*{\vec}[1]  {\boldsymbol{#1}}
\renewcommand*{\v}[1]  {\boldsymbol{#1}}
\newcommand* {\m}[1]  {\mathcal{#1}}
\newcommand*  {\s}[1]    {\mathsf{#1}}

\newcommand*  {\p}       {\partial}

\newcommand*  {\Vp}      {V_{\mathrm{pl}}}

\newcommand   {\sub}[2]  {{#1}_{\mathrm{#2}}}

\newcommand*  {\twovector}[2] {{\begin{pmatrix} $1 \\ $2 \end{pmatrix}}}

\renewcommand {\emph}[1]  {\textit{#1}}
\newcommand{\dd}{\mathrm{d}}

\shorttitle{Artificial precession}
\shortauthors{David M. Hernandez}

\begin{document}

\title{Artificial precession and instability in solar system and planetary simulations: analytic and numerical results}

\author[orcid=0000-0001-7648-0926,gname=David M.,sname=Hernandez]{David M. Hernandez}
\affiliation{Department of Physics, National Taiwan Normal University, Taipei 116, Taiwan}
\email[show]{dmhernandez@ntnu.edu.tw}

\begin{abstract}
Wisdom--Holman (WH) methods are algorithms used as a basis for a wide range of codes used to solve problems in solar system and planetary dynamics.  The problems range from the growth and migration of planets to the stability of the solar system.  In many cases, these codes work with Democratic Heliocentric Coordinates (DHC) which offer some advantages.  However, it has been noted these coordinates affect the dynamics of solar system bodies in simulations, in particular Mercury's, and introduce artificial precession which affects solar system stability.  In this work, we analytically derive the two-body artificial precession induced by DHC.  We show the effect is small for solar system bodies, but the artificial effect on Jupiter is $242$ times larger than on Mercury.  In a two-body Mercury-Sun system with general relativity (GR), artificial precession is negligible compared to GR precession, even with extreme timesteps that amplify the numerical effects.  A simple two-planet Mercury--Jupiter system without GR amplifies artificial precession significantly.  However, large artificial precession or artificial instability is not a danger unless one uses large timesteps that break the surrogate Hamiltonian approximation.    
\end{abstract}


\section{Introduction}
\label{sec:intro}

In studies over many dynamical timescales, planetary, planetesimal, and solar system dynamics usually use specialized algorithms based on the Wisdom--Holman (WH) method \citep{Kinoshitaetal91,WH91}.  These codes are used for studies ranging from stability analyses, to the growth of planets and planetary accretion, to the migration of planets.  The idea behind WH is simple: most planetary motion can be treated as a Keplerian two-body problem.  Perturbations from additional forces and planets are treated as a small effect, added in after two-body solutions.  

Democratic Heliocentric Coordinates (DHC) \citep{DLL98} are used in many such codes, like \texttt{WHFast512} \citep{Javaherietal2023}, \texttt{MERCURY} and \texttt{MERCURIUS} \citep{C99,Reinetal2019}, \texttt{SYMBA} \citep{Leeetal1997,DLL98,LD00}, \texttt{GENGA} \citep{Grimmetal2014}, \texttt{HNBODY} \citep{RauchHamilton2002}, and \texttt{TRACE}, \citep{HD2023,Luetal2024}.  Originally, WH was developed using Jacobi coordinates, which have advantages and disadvantages; for an overview see \cite{HD17}.  A disadvantage of DHC is that WH in DHC does not solve two-body problems exactly, in contrast to WH in Jacobi coordinates.  

\cite{Zeebe2015a} noted that the evolution of Mercury's eccentricity in simulations of the solar system depended on the coordinate system.  \cite{KaibRaymond2025} found the solar system was significantly more stable when using DHC.  \cite{Reinetal2026} have proposed an explanation for this stability in that WH in DHC introduces artificial precession to Mercury's orbit, in the same prograde direction as general relativity (GR), which pushes the $g_1$ secular frequency associated with Mercury ($= 559 ''/\mathrm{cent}$) further from $g_5 = 426''/\mathrm{cent}$, associated with Jupiter.  $g_1$ evolves diffusively or subdiffusively in time \citep{Abbotetal2024}, and solar system instabilities are associated with the case of $g_1$ approaching $g_5$.

In this paper, we derive and test the artificial two-body precession due to WH in DHC.  We find a small effect that is $242$ times stronger for Jupiter than for Mercury.  We also test two-body artificial precession in the presence of GR, finding a negligible effect, even for extreme timesteps.  Finally, we can match the qualitative \cite{Reinetal2026} strong artificial precession results with a simplified two-planet system. 

Section \ref{sec:notation} presents mathematical preliminaries.  In Section \ref{sec:prec} we derive and test two-body artificial precession.  Section \ref{sec:GR} studies two-body problems in the presence of GR.  We conclude in Section \ref{sec:conc}.
\section{Mathematical preliminaries}
\label{sec:notation}

A system of $N$ point particles subject to Newtonian gravitational forces has Hamiltonian, written in inertial Cartesian coordinates, 
\begin{equation}
H = T + V = \sum_{i=0}^{N-1} \frac{\vec{p}_i^2}{2m_i} - \sum_{ i=0}^{N-1} \sum_{j=i+1}^{N-1}\frac{Gm_im_j}{r_{i j}},
\label{eq:Hamilt}
\end{equation} \par\noindent 
\par\noindent
where we've identified the kinetic and potential energies, $T$ and $V$.  $G$ is Newton's gravitational constant, $\v{p}_i$ is the non-relativistic momentum of particle $i$, $m_i$ its mass, and $\v{r}_i$ its position. $\v{r}_{ij} = \v{r}_i - \v{r}_j$.  Angular momentum $\v{L} = \sum_i \v{r}_i \times \v{p}_i$ is conserved by Hamiltonian \eqref{eq:Hamilt}.

Given a phase space state of a system of particles described by \eqref{eq:Hamilt}, our goal is to calculate the state at a time $t$.  Assume the state at $t = 0$ is $\v{z}(0) = (\v{r}(0),\v{p}(0))$.  Hamilton's equations are difficult to solve:
\begin{equation}
\begin{aligned}
\dot{\v{p}}_i &= -G \sum_{j \ne i} \frac{m_i m_j}{r_{ij}^3} \v{r}_{ij}, \\
\dot{\v{r}}_i & = \frac{\v{p}_i}{m_i}.
\label{eq:Hamilteq}
\end{aligned}
\end{equation} \par\noindent
\par\noindent
To arrive at the same result, we can use a different mathematical formulation.  Given a function $F(\v{z})$, define an operator $\hat{F} \v{z} = \{\v{z}, F\}$, where $\{\}$ are Poisson brackets:
\begin{equation}
\label{eq:poisson}
\{K, F\} = \sum_i \left(\frac{\partial K}{\partial \v{r}_i} \cdot \frac{\partial F}{\partial \v{p}_i} - \frac{\partial K}{\partial \v{p}_i} \cdot \frac{\partial F}{\partial \v{r}_i} \right),
\end{equation} \par\noindent
\par\noindent
with $K$ a second function of $\v{z}$.  Any canonical basis, not just Cartesian coordinates, can be used to compute derivatives in Eq. \eqref{eq:poisson}.  Hamilton's equations are then,
\begin{equation}
\label{eq:dotdef}
\dot{\v{z}} = \{\v{z}, H\}.
\end{equation} \par\noindent
\par\noindent
It follows that $F = \{F,H\}$.  The solution to Hamilton's equations is given by the infinite series,
\begin{equation}
\label{eq:taylorexp}
\v{z}(t) =  \exp\left( t \hat{H} \right) \v{z}(0) = \sum_{i = 0}^{\infty} \frac{t^i}{i!} \hat{H}^i \v{z}(0).
\end{equation} \par\noindent

Obviously, solution \eqref{eq:taylorexp} is not very useful or practical.  A simpler, but approximate, solution path is often taken in dynamical systems work.  We separate Hamiltonian \eqref{eq:Hamilt} into two or more pieces that are easier to solve and often integrable.  This approach yields what is known as a symplectic integrator.  If the masses in \eqref{eq:Hamilt} are not too different in scale, some approaches are given by \cite{GBP14,HB15,DH17}.  In this paper, however, we specialize to the case when $m_0 \gg m_{i {\scriptscriptstyle >} 0}$.  We call $m_0$ the ``star'' and $m_{i {\scriptscriptstyle >} 0}$ ``planets.''    As an example of such a splitting, let $H = \sub{A}{C} + \sub{B}{C}$, with,
\begin{equation} 
\label{eqs:WHI:A+B}
\begin{aligned}
	\sub{A}{C}  &=
		\sum_{i\neq0}\frac{\vec{p}_i^2}{2m_i} - \frac{Gm_0m_i}{r_{i 0}}
	\qquad\text{and}\qquad \\
	\sub{B}{C}  &= \frac{\vec{p}_0^2}{2m_0}
		- \sum_{0{\scriptscriptstyle <}i{\scriptscriptstyle <}j}\frac{Gm_im_{\!j}}{r_{i j}}. \\
\end{aligned}
\end{equation} \par\noindent
$\sub{A}{C}$ is a sum of $n$ independent Kepler two-body problems, with $n = N-1$ the number of planets.  Extensive tools are available to solve $\sub{A}{C}$.  $\sub{B}{C}$ is a trivial integrable Hamiltonian.  Solving $\sub{B}{C}$ maps state $\v{z}$ at $t = 0$ to $\v{z}^\prime$ at $t$:
\begin{equation}
\label{eq:mapc}
\begin{aligned}
\v{p}_{i {\scriptscriptstyle >} 0}^{\prime} &= \v{p}_{\scriptstyle i {\scriptscriptstyle >} 0} - t \sum_{j \ne i, j {\scriptscriptstyle >} 0} \frac{G m_i m_j}{r_{ij}^3} \v{r}_{i j}, \\
\v{r}_0^{\prime} &= \v{r}_0 + t \frac{\v{p}_0}{m_0}.
\end{aligned}
\end{equation} \par\noindent
In a barycentric frame, $\sub{B}{C}/\sub{A}{C} = \mathcal O(\epsilon)$, with 
\begin{equation}
\label{eq:eps}
\epsilon = \max\ ( m_i/m_0 ),
\end{equation} \par\noindent 
a small parameter.  To see this, use the scaling arguments, $v_i^2 \propto m_0/r_{i0}$ and $p_0^2 \propto m_i^2 v_i^2$, with $\v{v}_i = \dot{\v{r}}_i$.  

Having effectuated our Hamiltonian splitting \eqref{eqs:WHI:A+B}, rather than using Eq. \eqref{eq:taylorexp}, we write the approximate solution, 
\begin{equation}
\v{z} (t) = \exp (t \sub{A}{C}) \exp (t  \sub{B}{C}) \v{z}(0) = \exp(t \hat{\tilde{H}}_{\mathrm{Eul}}) \v{z}(0),
\label{eq:Eul}
\end{equation} \par\noindent
which is only a good approximation for $t$ small.  $\tilde{H}_{\mathrm{Eul}}$ is a surrogate Hamiltonian, close to $H$ for small $t$:
\begin{equation}
\tilde{H}_{\mathrm{Eul}} = H + t H_1 + t^2 H_2 + \hdots,
\label{eq:HEul}
\end{equation} \par\noindent
where the $H_i$ can be computed via the Baker--Campbell--Hausdorff (BCH) formula \citep{Campbell1996,Campbell1997,Baker1902b,Baker1905,Hausdorff1906,Dynkin1947,hair06}.  

We can do better still than Eq. \eqref{eq:Eul} in our approximation to $\v{z}(t)$.  We first find a better canonical coordinate system that reduces the number of degrees of freedom.  A suitable choice is DHC, written $(\v{Q},\v{P})$.  $\v{Q}_0$ and $\v{P}_0$ are the center of mass and total momentum, respectively, while $\v{Q}_{i {\scriptscriptstyle >} 0}$ and $\v{P}_{i {\scriptscriptstyle >} 0}$ are heliocentric positions and barycentric momenta, respectively.  We obtain these coordinates via linear transformations,
\begin{subequations}
	\label{eqs:DHC:Q,P}
\begin{align}
	\vec{Q}_0 &= \sum_j\frac{m_j}{M}\vec{r}_{\!j},
	\\
	\vec{Q}_{i\neq0} &= \vec{r}_i-\vec{r}_0,
	\\
	\vec{P}_0 &= \sum_j\vec{p}_{\!j},
	\\
	\vec{P}_{i\neq0} &= \vec{p}_i - \frac{m_i}{M}\vec{P}_0 = 
		\vec{p}_i - \frac{m_i}{M}\sum_j\vec{p}_{\!j},
\end{align}
\end{subequations} \par\noindent
with $M = \sum_i m_i$.  In DHC, working in a barycentric frame, we can write,
\begin{equation} 
\label{eq:DHC:T}
\begin{aligned}
	H &= 
	  \underbrace{\sum_{i\neq0} \frac{\vec{P}_i^2}{2m_i}}_{T_1}
	  + \underbrace{\frac{1}{2m_0} \left(\sum_{i\neq0}\vec{P}_i\right)^2}_{T_0} - \underbrace{\sum_{i\neq0} \frac{Gm_im_0}{|\vec{Q}_i|}}_{V_{\odot}} \\
	    &- \underbrace{\sum_{0{\scriptscriptstyle <}i{\scriptscriptstyle <}j} \frac{Gm_im_{\!j}}{|\vec{Q}_{i}-\vec{Q}_{\!j}|}}_{\sub{V}{pl}},
	    \end{aligned}
\end{equation} \par\noindent
where we identified functions $T_1$, $T_0$, $V_\odot$, and $\sub{V}{pl}$, equal to the planetary kinetic energies, solar kinetic energy, potential energy from Sun-planet interactions, and potential energy from planet-planet interactions, respectively.  It is easy to show,
\begin{equation}
\label{eq:Angmom}
\v{L} = \sum_i \v{r}_i \times \v{p}_i = \sum_i \v{Q}_i \times \v{P}_i,
\end{equation} \par\noindent
and,
\begin{equation}
\label{eq:Lbrack}
\{\v{L},T_1\} = \{\v{L},T_0\} = \{\v{L},V_\odot \} = \{\v{L},\sub{V}{pl}\} = 0:
\end{equation} \par\noindent
all the functions conserve $\v{L}$.

The next improvement comes from increasing the order of the error in Eq. \eqref{eq:HEul} from $t^1$ to $t^2$.  Now, using a small timestep $h$, we can write as solution,
\begin{equation}
\begin{aligned}
\v{z} (h) &= \exp \left(\frac{h}{2} \sub{\hat{A}}{D}\right) \exp \left(h \sub{\hat{B}}{D} \right) \exp \left( \frac{h}{2} \sub{\hat{A}}{D} \right) \v{z}(0) \\
 &= \exp \left(h \hat{\tilde{H}}_{\mathrm{ABA}} \right) \v{z}(0),
\label{eq:leap}
\end{aligned}
\end{equation} \par\noindent
with,
\begin{equation}
\label{eq:AB}
	\sub{A}{D}  = T_1 + V_\odot
	\qquad\text{and}\qquad
	\sub{B}{D}  = T_0 + \sub{V}{pl}.
\end{equation} \par\noindent
If $t = mh$, $\v{z}(t)$ is then found by iterating Eq. \eqref{eq:leap} for $m$ steps.  We can show again that $\sub{B}{D}/\sub{A}{D} = \mathcal O(\epsilon)$.  Eq. \eqref{eq:leap} is the ``ABA'' second order method; just as valid, we can construct a ``BAB'' method by swapping $\sub{A}{D}$ and $\sub{B}{D}$.  $\sub{\tilde{H}}{BAB}$ and $\sub{\tilde{H}}{ABA}$ are closer to $H$ than $\tilde{H}_{\mathrm{Eul}}$:
\begin{align}
	\label{eq:WH:Herr:BAB}
	\sub{\tilde{H}}{BAB} &= H
	+\frac{h^2}{12}\{\{\sub{B}{D},\sub{A}{D}\},\sub{A}{D}\} \\
	& -\frac{h^2}{24}\{\{\sub{A}{D},\sub{B}{D}\},\sub{B}{D}\} 
	+\mathcal{O}(h^4), \notag
	\\
	\label{eq:WH:Herr:ABA}
	\sub{\tilde{H}}{ABA}  &= H
	-\frac{h^2}{24}\{\{\sub{B}{D},\sub{A}{D}\},\sub{A}{D}\} \\
	& +\frac{h^2}{12}\{\{\sub{A}{D},\sub{B}{D}\},\sub{B}{D}\} 
	+\mathcal{O}(h^4). \notag
\end{align} \par\noindent

If we let $\sub{A}{D} \propto \epsilon^0$, then $\sub{B}{D} \propto \epsilon^1$ and $\v{P}_{i \ne 0} \propto \epsilon^0$, so that a Poisson bracket does not change the order of $\epsilon$.  This leads us to find the leading order error for Eqs. \eqref{eq:WH:Herr:BAB} and \eqref{eq:WH:Herr:ABA} is $\mathcal O(\epsilon h^2)$.  We can write the error terms in more detail:
\begin{subequations} \label{eq:err:WHI}
\begin{align}
	\{\{\sub{A}{D},\sub{B}{D}\},\sub{B}{D}\}
		&= \{\{V_\odot,T_0\},T_0\} + \{\{T_1,\Vp\},\Vp\}, \text{  and}\\
	\{\{\sub{B}{D},\sub{A}{D}\},\sub{A}{D}\}
		&= \{\{T_0,V_\odot\},V_\odot\} + \{\{\Vp,T_1\},T_1\}
		\\
		&-	\{\{T_1,\Vp\},V_\odot\} - \{\{V_\odot,T_0\},T_1\}. \notag
\end{align}
\end{subequations} \par\noindent
The explicit forms for Eqs. \eqref{eq:err:WHI} in terms of $(\v{Q},\v{P})$ are presented in \cite[][Eqs. 41]{HD17}.  Both methods ABA and BAB form the basis of the list of codes in Section \ref{sec:intro}.  

\section{Two-body precession}
\label{sec:prec}
Now, consider the case of two bodies, when $\sub{V}{pl} = 0$.  Eqs. \eqref{eq:err:WHI} simplify:
\begin{subequations} \label{eq:err:simple}
\begin{align}
	\{\{\sub{A}{D},\sub{B}{D}\},\sub{B}{D}\}
		&= \{\{V_\odot,T_0\},T_0\}, 
	\qquad\text{and}\\
	\{\{\sub{B}{D},\sub{A}{D}\},\sub{A}{D}\}
		&= \{\{T_0,V_\odot\},V_\odot\}
		- \{\{V_\odot,T_0\},T_1\}.
\end{align}
\end{subequations} \par\noindent
The fact that $\sub{\tilde{H}}{BAB}$ and $\sub{\tilde{H}}{ABA}$ are not exactly equal to $H$ implies the two-body problem is not solved exactly.  However, $\sub{\tilde{H}}{BAB}$ and $\sub{\tilde{H}}{ABA}$ are to excellent approximation conserved for $h$ small,\footnote{More precisely, their conservation is a function of the phase along the orbit.} and $\v{L}$ is exactly conserved by the maps per Eq. \eqref{eq:Lbrack}, regardless of $h$.  Thus, for small $h$, we have a set of three integrals of motion in involution with each other; for example   $\sub{\tilde{H}}{BAB}$, $L^2$, and $\sub{L}{z}$.  Thus, the motion is integrable.  But there is no conservation of the Runge--Lenz vector, so the orbit does not close.  We can say that the motion is integrable, but not superintegrable \citep{Fasso2005}. 

We can calculate, say, Eq. \eqref{eq:WH:Herr:BAB} explicitly, using Eqs. 41 from \cite{HD17}.  There are only $3$ degrees of freedom, $(\v{Q}_1, \v{P}_1)$, which we denote simply as $(\v{Q},\v{P})$:
\begin{equation}
\begin{aligned}
\sub{\tilde{H}}{BAB} &= 
{\frac{P^2}{2 \mu} - \frac{G m_0 m_1}{Q}}- \frac{G h^2 \epsilon }{24} \left( \frac{P^2}{Q^3} - \frac{3 (\v{P} \cdot \v{Q})^2}{Q^5} \right) \\
&+ \frac{G h^2}{12} \left( \frac{G m_0 m_1^2}{Q^4} + \frac{P^2}{Q^3} - \frac{3 (\v{P} \cdot \v{Q})^2}{Q^3} \right) + \mathcal O(h^4),
\label{eq:Htild2bod}
\end{aligned}
\end{equation} \par\noindent
where $\mu = m_0 m_1/(m_0 + m_1)$ is called the reduced mass, and 
\begin{equation}
\label{eq:eps2bod}
\epsilon = m_1/m_0, 
\end{equation} \par\noindent
from Eq. \eqref{eq:eps}.

In a barycentric frame, Eq. \eqref{eq:Hamilt} is,
\begin{equation}
H = \frac{p_r^2}{2 \mu} - \frac{G m_0 m_1}{r_{01}} = \frac{P^2}{2 \mu} - \frac{G m_0 m_1}{Q},
\label{eq:H2bod}
\end{equation} \par\noindent
where $\v{p}_r = \mu (\v{v}_1 - \v{v}_0) = \mu ( \v{P}/m_1 + \v{P}/m_0) = \v{P}$.  The Runge--Lenz (RL) vector is,
\begin{equation}
\v{R} = \v{P} \times \v{L} - c_0 \frac{\v{Q}}{Q},
\end{equation} \par\noindent
with $c_0 = G m_0^2 m_1^2 /M$.  $\v{R}$ points from the barycenter to periapse and 
\begin{equation}
\label{eq:Rcons}
\{\v{R},H\} = 0.  
\end{equation} \par\noindent
Instantaneous and average precession rates are constructed:
\begin{equation}
\label{eq:preceq}
\dot{\varpi}= \frac{ \left(\v{R} \times \dot{\v{R}} \right)_{z}}{R^2} \qquad \text{and} \qquad
\langle \dot{\varpi} \rangle= \left \langle \frac{ \left(\v{R} \times \dot{\v{R}} \right)_{z}}{R^2} \right \rangle,  
\end{equation} \par\noindent
where a $z$ subscript indicates we take the $z$-component, and $\hat{z}$ is defined as $\hat{\v{L}}$.  Time averages are carried out by integrating over the orbit defined by $H$, which differs from the BAB orbit by $\mathcal O(h^2)$, per Eq. \eqref{eq:Htild2bod}.  This correction has no leading order impact on our results.  Thus, our numerical BAB orbit yields,
\begin{equation}
\label{eq:approx}
R^2 =   c_0^2 e^2 + \mathcal O(h^2) \qquad\text{and}\qquad \v{R} = \v{R}_0 + \mathcal O(h^2),
\end{equation} \par\noindent
with $e$ and $\v{R}_0$ the eccentricity and invariant RL vector described by $H$.  The difference between the BAB and $H$ semi-major axes, $\sub{a}{BAB} - a$, with $\sub{a}{BAB}$ an osculating orbital element, remains small when $h$ is small, over exponentially long times \citep[][Sec. IX.8, Thm. 8.1]{hair06}; this is a fundamental property of symplectic methods \citep{chan90,SC93,leim04,blanescasas2017,Tremaine2023}.  Moreover, $\v{L}$ is exactly conserved by $H$ and $\sub{\tilde{H}}{BAB}$.  It can be written,
\begin{equation}
\label{eq:Ldef}
{L} = \mu \sqrt{G M a (1 - e^2 ) } = \mu \sqrt{G M \sub{a}{BAB} (1 - \sub{e}{BAB}^2 ) },
\end{equation} \par\noindent
which implies $e - \sub{e}{BAB}$ also cannot drift secularly in time.  This can be numerically verified.  Thus, we have shown the near constancy of $\sub{a}{BAB} $ and $\sub{e}{BAB}$, but no similar constraint exists on the argument of pericenter, $\varpi$.  This causes a precessing orbit in general.

Next,
\begin{equation}
\dot{\v{R}} = \{\v{R}, \sub{\tilde{H}}{BAB}\} = \{\v{R}, \sub{H}{err}\},
\end{equation} \par\noindent
where $\sub{H}{err} = \sub{\tilde{H}}{BAB} - H$ is the error Hamiltonian.  
With manipulation of Eqs. \eqref{eq:WH:Herr:BAB} and \eqref{eq:err:simple}, we can write, 
\begin{equation}
\sub{H}{err} = \underbrace{\frac{h^2}{24} \left[ \{ \{ V_{\odot}, T_0 \}, H \} \right]}_{\sub{H}{A}} 
+ \underbrace{\frac{h^2}{24} \left[ \{ \{ V_\odot, T_0 \}, T_0 \} \right]  +\mathcal O(h^4)}_{\sub{H}{B}},
\end{equation} \par\noindent 
where we have identified $\sub{H}{A}$ and $\sub{H}{B}$.  The Jacobi identity for Poisson brackets, for $F(\v{z})$, $J(\v{z})$, and $K(\v{z})$ is,
\begin{equation}
\label{eq:Jacobiident}
\{F, \{J,K\} \} + \{J, \{K,F\} \} + \{K, \{F,J\} \} = 0.
\end{equation} \par\noindent
Using it, and Eq. \eqref{eq:Rcons}, we find,
\begin{equation}
\begin{aligned}
\label{eq:manip}
\{\v{R}, \sub{H}{A} \} &= \{ \v{R}, \{S, H\} \} = -\{S, \{H, \v{R} \} \} - \{H, \{\v{R}, S\} \} \\
&= \frac{\dd}{\dd t} \left( \{\v{R}, S \} \right),
\end{aligned}
\end{equation} \par\noindent
where $S = (h^2/24) \{V_{\odot}, T_0\}$ and the time derivative is along the orbit described by $H$.  

To leading order in $h$, $\v{R} \times \dot{\v{R}} = \v{R}_0 \times \dot{\v{R}}$ and the average precession over an orbit contributed by Eq. \eqref{eq:manip} must be $0$ as the orbit is periodic.  Thus, only $\sub{H}{B}$ contributes to precession:
\begin{equation}
\begin{aligned}
\label{eq:Rdotcomm}
\dot{\v{R}} = \{ \v{R}, \sub{H}{B} \} &= \frac{G h^2 \epsilon}{24} \left\{ \v{P} \times \v{L} - c_0 \frac{\v{Q}}{Q},   \frac{P^2}{Q^3} - \frac{3 (\v{P} \cdot \v{Q})^2}{Q^5}  \right\} \\
&+ \mathcal O(h^4).
\end{aligned}
\end{equation} \par\noindent
Because of \eqref{eq:Lbrack}, $\v{L}$ can be taken outside all Poisson brackets.  The result of Eq. \eqref{eq:Rdotcomm} is straightforward:
\begin{equation}
\begin{aligned}
\dot{\v{R}} &= \frac{G h^2 \epsilon}{24} \left[ \left(\v{Q} \times \v{L} \right) \left( \frac{3 P^2}{Q^5} -  \frac{15(\v{P} \cdot \v{Q})^2}{Q^7} + \frac{4 c_0}{Q^6}\right)  \right. \\
& \left. + \left( \v{P} \times \v{L} \right) \frac{6 (\v{P} \cdot \v{Q})}{Q^5} - \frac{2 c_0 \v{P}}{Q^4} + \frac{2 c_0 (\v{P} \cdot \v{Q}) \v{Q}}{Q^6} 
\right] +\mathcal O(h^4).
\end{aligned}
\end{equation} \par\noindent  
We then compute the cross product, and take the $z$ component:
\begin{equation}
\label{eq:Rdot}
\begin{aligned}
(\v{R} \times \dot{\v{R}} )_z &= \frac{G h^2 \epsilon L}{24} \\
&\times \left[ \left( -L^2 + c_0 Q \right) \left( \frac{3 P^2}{Q^5} 
- \frac{15 (\v{P} \cdot \v{Q})^2 }{Q^7} + \frac{4 c_0}{Q^6} \right) \right. \\
& \left. +\frac{8 c_0 (\v{P} \cdot \v{Q})^2}{Q^6} - \frac{2 c_0}{Q^4} \left( P^2 - \frac{c_0}{Q} \right)
\right] +\mathcal O(h^4),
\end{aligned}
\end{equation} \par\noindent
Using Eqs. \eqref{eq:preceq}, \eqref{eq:approx}, and \eqref{eq:Ldef}, the total precession is,
\begin{equation}
\label{eq:crossp}
\begin{aligned}
\dot{\varpi}_{\mathrm{tot}} &= \frac{G h^2 \epsilon \sqrt{c_0 a (1 - e^2)}}{24 e^2} \\
& \times \left[ 
\left( -a (1-e^2) + Q \right) \left( \frac{3 v^2}{G M Q^5}  
- \frac{15(\v{v} \cdot \v{Q})^2}{GM Q^7} + \frac{4}{Q^6} \right) \right. \\
& \left. + \frac{8(\v{v} \cdot \v{Q})^2}{GM Q^6} - \frac{2}{Q^4} \left( \frac{v^2}{GM} - \frac{1}{Q} \right)
\right] + \mathcal O(h^4),
\end{aligned}
\end{equation} \par\noindent
where $\v{v} = \v{P}/\mu$ is the relative velocity.  

Next, we must compute the average over a period for Eq. \eqref{eq:crossp}.  Writing the expressions for $\v{v}(t)$ and $\v{Q}(t)$ for a two-body orbit, and converting to an average over true anomaly, the following time averages are computed:
\begin{subequations}
	\label{eqs:average}
\begin{align}
\left \langle \frac{v^2}{Q^5} \right \rangle 
&= \frac{G M}{a^6} \frac{4 + 22e^2 + 9e^4}{4 (1 - e^2)^{9/2}},
\\
\left \langle \frac{v^2}{Q^4} \right \rangle
&= \frac{G M}{a^5} \frac{2 + 7e^2 + e^4}{2 (1 - e^2)^{7/2}},
\\
\left \langle \frac{(\v{v} \cdot \v{Q})^2 }{Q^7} \right \rangle
&= \frac{G M e^2}{a^6} \frac{4 + 3e^2}{8 (1 - e^2)^{9/2}},
\\
\left \langle \frac{(\v{v} \cdot \v{Q})^2 }{Q^6} \right \rangle
&= \frac{G M e^2}{a^5} \frac{4 + e^2}{8 (1 - e^2)^{7/2}},
\\
\left \langle \frac{1}{Q^6} \right \rangle
&= \frac{1}{a^6} \frac{8 + 24e^2 + 3e^4}{8(1 - e^2)^{9/2}},
\\
\left \langle \frac{1}{Q^5} \right \rangle
&= \frac{1}{a^5} \frac{2 + 3e^2}{2(1 - e^2)^{7/2}}.
\end{align}
\end{subequations} \par\noindent
Taking an average of Eq. \eqref{eq:crossp} is then,
\begin{equation}
\left \langle \dot{\varpi}_\mathrm{tot} \right \rangle =
 -\frac{h^2}{4} \epsilon \left( \frac{m_0 m_1}{M^2} \right) \left( \frac{G M}{a^3} \right) ^{3/2} \frac{1 + \frac{e^2}{4}}{(1-e^2)^3} + \mathcal O(h^4),\\
\end{equation} \par\noindent
or, preferring $m_0$ rather than $M$, and keeping the lowest order term,
\begin{equation}
\label{eq:omfin}
\left \langle \dot{\varpi} \right \rangle =-\frac{h^2}{4} \epsilon^2 \omega_0^3  \frac{1 + \frac{e^2}{4}}{(1-e^2)^3} ,
\end{equation} \par\noindent
with $\omega_0 = (G m_0/a^3)^{1/2}$ and $\epsilon$ from Eq. \eqref{eq:eps2bod}.  This expression has error $ \mathcal O(\epsilon^3 h^2 + h^4)$ compared to $\langle \dot{\varpi}_{\mathrm{tot}} \rangle$.  Note the negative sign, indicating the precession is always retrograde.  The artificial precession goes to $0$ as the timestep is decreased to $0$.  It also grows with $\omega_0$.  As $e \rightarrow 1$, $\left \langle \dot{\varpi} \right \rangle$ increases sharply.  At extreme eccentricities we observe our derivation and analytic prediction break down.  However, if one uses timesteps ``resolving'' pericenter given $e$, this breakdown does not occur in our tests.

\subsection{What about ABA?}
\label{sec:ABA}
We now calculate the precession frequency for the dual map ABA.  Letting $\Delta H = \sub{\tilde{H}}{ABA} - \sub{\tilde{H}}{BAB}$, Eqs. \eqref{eq:WH:Herr:BAB} and \eqref{eq:WH:Herr:ABA} say,
\begin{equation}
\Delta H = \frac{h^2}{8} \{\{\sub{A}{D},\sub{B}{D}\},H\} + \mathcal O(h^4).
\end{equation} \par\noindent
Defining, $\Delta \dot{\v{R}} = \dot{\v{R}}_{\mathrm{ABA}} - \dot{\v{R}}_{\mathrm{BAB}}$, we get, using Eqs. \eqref{eq:Jacobiident} and \eqref{eq:Rcons}.
\begin{equation}
\begin{aligned}
\Delta \dot{\v{R}} &= \frac{h^2}{8}\{\v{R}, \{\{\sub{A}{D},\sub{B}{D}\},H\} \} + \mathcal O(h^4)\\
&= \frac{h^2}{8}\{H, \{ \{\sub{A}{D},\sub{B}{D}\},\v{R} \} \} + \mathcal O(h^4) \\
&= -\frac{h^2}{8} \frac{\dd}{\dd t} \{ \{\sub{A}{D}, \sub{B}{D}\}, \v{R} \} + \mathcal O(h^4).
\label{eq:manip2}
\end{aligned}
\end{equation} \par\noindent
Using the same logic as that under Eq. \eqref{eq:manip}, we see, 
\begin{equation}
\left \langle \dot{\v{R}}_{\mathrm{ABA}} \right \rangle = \left \langle \dot{\v{R}}_{\mathrm{BAB}} \right \rangle + \mathcal O(h^4), 
\end{equation} \par\noindent
so that expression \eqref{eq:omfin} is valid for ABA as well.
\subsection{Numerical verification}
\label{sec:numver}
To verify Eq. \eqref{eq:omfin}, we set up a two-body system, initialized at apocenter, with $\v{r}_{10} = a (1+e) \hat{x}$ and $\dot{\v{r}}_{10} = \sqrt{{G M (1-e)}/({a (1+e)})} \hat{y}$ so that $\hat{\v{{R}}}(0) = -\hat{x}$.  We use BAB to solve for the motion in time.  In units of solar mass, au, and yr, we set $m_0 = 1$, $m_1 = 0.001$, $a=1$.  $e$ is chosen as $25$ linearly spaced values between $0.1$ and $0.8$.  $h = P/100$, where $P = 1.00$ is the period.  We run for $t_{\mathrm{max}} = 10^4 P$.  The angle $\theta$ of $\v{R}$ with ${\v{R}}(0)$ is calculated at every step, with the orientation of $\theta$ determined by $\hat{\v{L}}$.  Numerically, $\theta$ is found to decrease linearly in time over $t_{\mathrm{max}}$; however, on short timescales there are departures from linearity due to numerical artifacts of the timestep interacting with timescales in the system; such numerical artifacts are well-known and average out over longer timescales.  Thus, we define 
\begin{equation}
\left \langle \dot{\varpi}_{\mathrm{num}} \right \rangle = \Delta \theta/t_{\mathrm{max}}.  
\end{equation} \par\noindent

\begin{figure}
	\includegraphics[width=80mm]{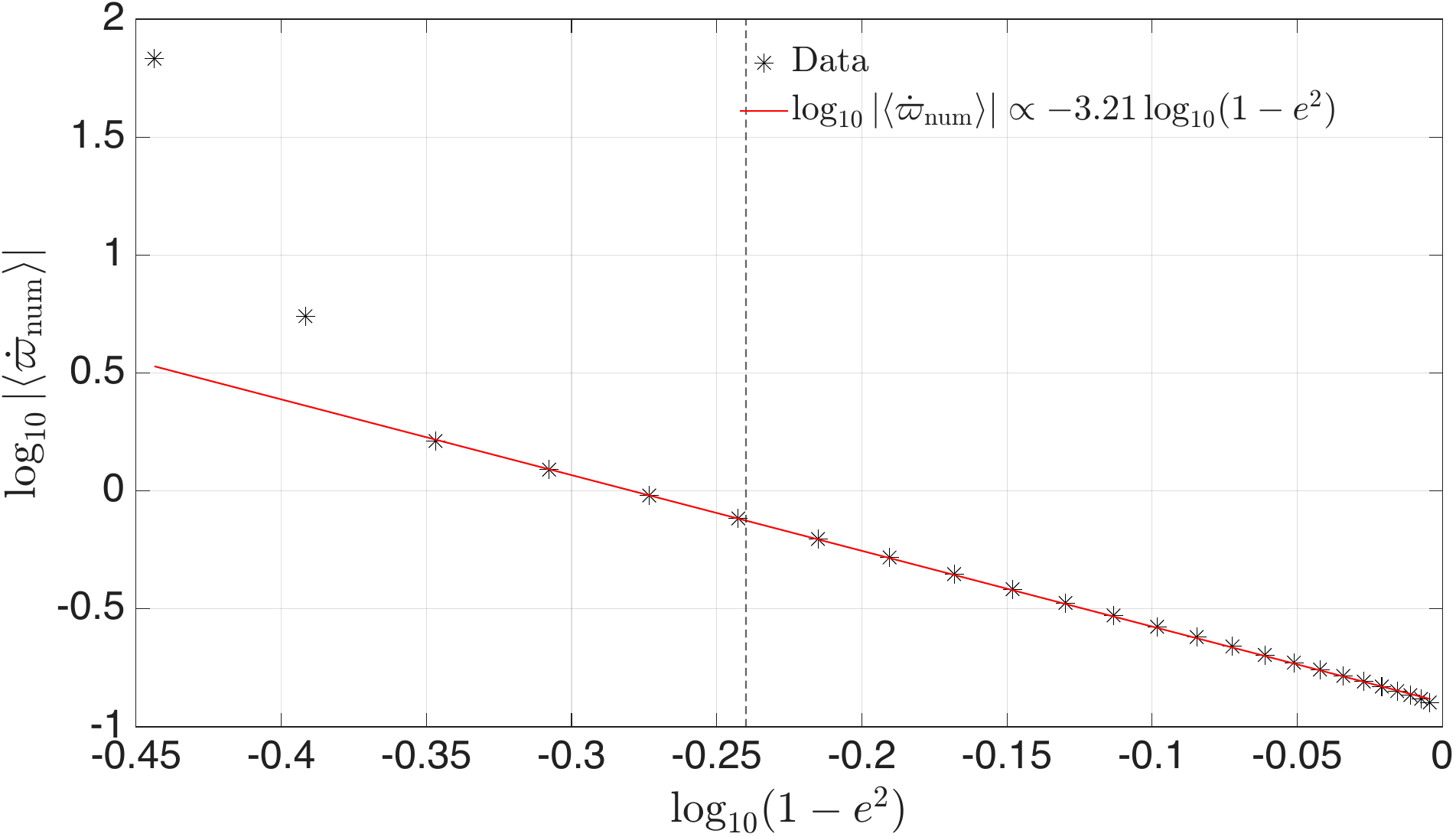}
	\caption{
	Numerical verification of the dependence on eccentricity of the two-body precession rate, Eq. \eqref{eq:omfin}.  In units of solar mass, au, and yr, we let $m_0 = 1$, $m_1 = 0.001$, $a = 1$, $h = P/100$, $P = 1.00$, and $t_{\mathrm{max}} = 10^4 P$, where $t_{\mathrm{max}} $ is the runtime.  The expected slope of $\log_{10} (1-e^2)$ versus $\log_{10} \left \langle \dot{\varpi}_{\mathrm{num}} \right \rangle$ is $s \in [-3.25, -3]$ as $e$ is increased to $1$.  At large $e$, $e > 0.74$, our precession derivation breaks down.  A least squares $s = -3.21$ for $e \le 0.74$ is plotted, and is consistent with theory.  The dashed line indicates a lower bound of the maximum resolved eccentricity in the context of planetary systems.
	\label{fig:escale}
  	}
\end{figure}

As predicted by \eqref{eq:omfin}, we find the precession to be retrograde, regardless of the orientation of $\hat{\v{L}}$.  We check the scaling of \eqref{eq:omfin} with $e$, written $g = (1+e^2/4)/(1-e^2)^3$.  Let $x = 1-e^2$.  Then, we find the slope, 
\begin{equation}
s = \frac{\dd \log g}{\dd \log x} = -3 - \frac{x/4}{1 + \frac{1}{4}(1-x)},
\end{equation} \par\noindent
so that $s \in [-3.25, -3]$ as $e$ increases from $0$ to $1$.  Figure \ref{fig:escale} plots $\log x$ versus $\log \left \langle \dot{\varpi}_{\mathrm{num}} \right \rangle$. 
For the data excluding the smallest two $x$ values, $e \le 0.74$, a least squares fit finds $s = -3.21$.  Our $x$ distribution is weighted towards $1$ ($\log_{10} x = 0$), so an $s$ approaching $-3.25$ agrees with theory.  As $e \rightarrow 0$, $\left \langle \dot{\varpi} \right \rangle =  -h^2 \epsilon^2 \omega_0^3/4 = - 0.13 ''/\mathrm{cent}$.

For $e > 0.74$ (two smallest $x$ points of Fig. \ref{fig:escale}), there is a clear change of slope, which we verified by testing smaller $x$ than those shown, although the scaling with $h^2$ remains.  However, agreement with \eqref{eq:omfin} is reestablished if we reduce $h$ further.  Changing the initial phase to pericenter also improves agreement with theory up to $e = 0.80$.  At high $e$, it's clear the derivation leading up to Eq. \eqref{eq:omfin} no longer holds.

In the context of planetary systems with additional planets, \cite{Hernandezetal2020} deduce a lower bound on the timestep required to resolve the pericenter timescale, $\tau_{\dot{f}}$, as $h > h_0 = \tau_{\dot{f}}/16$.  For our test, the corresponding $e$ for $h_0$ is $e_0 = 0.65$ and is shown as a dashed line.  The theory is consistent with the numerical results for $e < e_0$.

\begin{figure}
	\includegraphics[width=80mm]{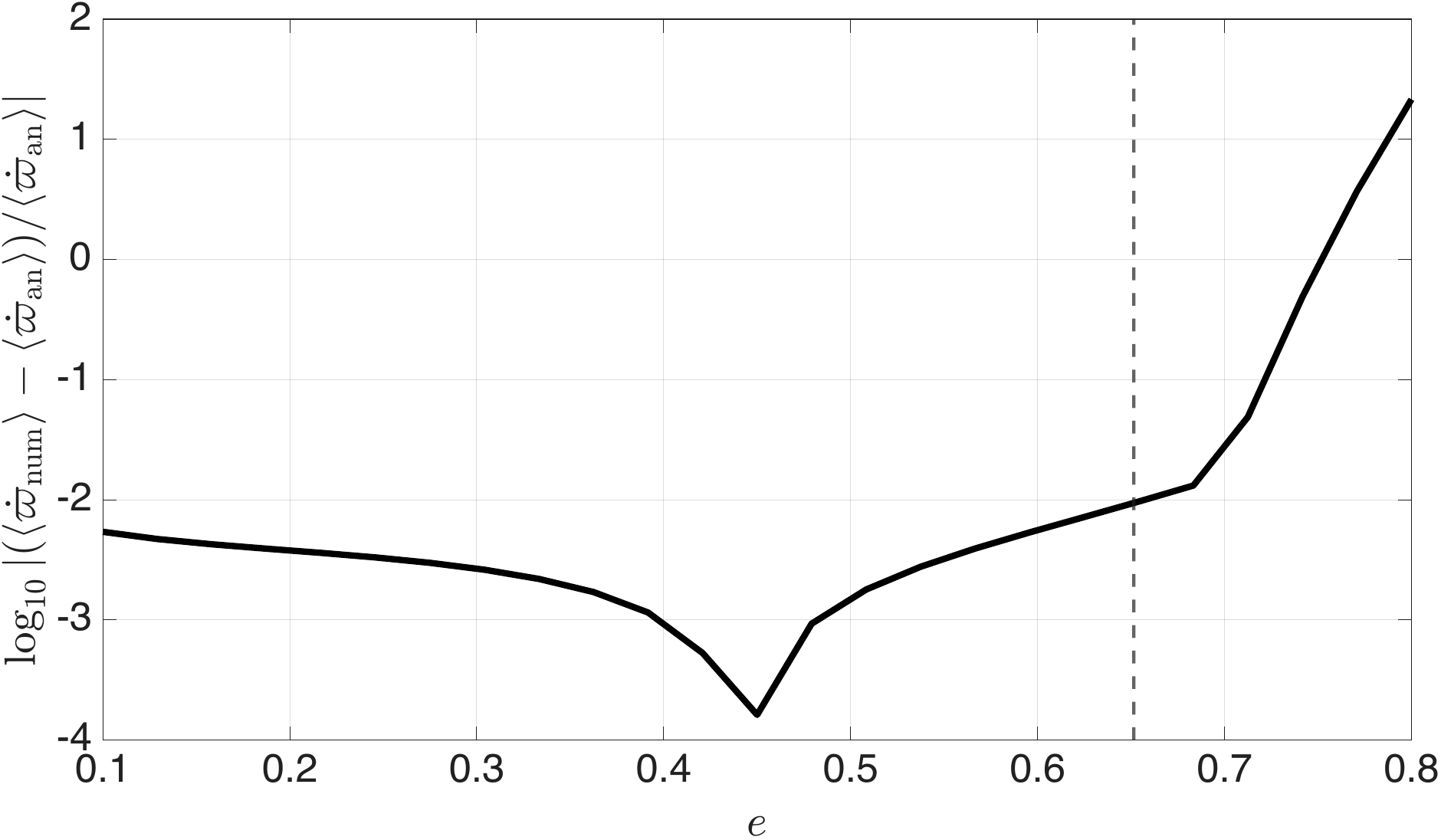}
	\caption{ The relative error between the predicted ($\left \langle \dot{\varpi}_{\mathrm{an}} \right \rangle$) and numerical $(\left \langle \dot{\varpi}_{\mathrm{num}} \right \rangle)$ precession rate as a function of $e$.  The system and parameters are the same as those for Fig. \ref{fig:escale} but the runtime has been extended to $10^5 P$.  While Fig. \ref{fig:escale} tests the scaling with $e$, here we test the full Eq. \eqref{eq:omfin}.  The agreement is good for $e < 0.74$.  The dashed line indicates a lower bound of the maximum resolved eccentricity in the context of planetary systems.
	\label{fig:omerr}
  	}
\end{figure}

Fig. \ref{fig:omerr} plots the relative error between the predicted precession ($\left \langle \dot{\varpi}_{\mathrm{an}} \right \rangle$) and $\left \langle \dot{\varpi}_{\mathrm{num}} \right \rangle$ as a function of $e$, but now for runtime $10^5 P$.  This plot tests the accuracy of the coefficients besides $g$ in Eq. \eqref{eq:omfin}.  The agreement is good for $e < 0.74$ (which includes $e < e_0$).  Note the minimum in the relative error at $e_1 \approx 0.45$: this corresponds to a change in sign of $\left \langle \dot{\varpi}_{\mathrm{an}} \right \rangle - \left \langle \dot{\varpi}_{\mathrm{num}} \right \rangle$.  For $e < e_1$, the analytic estimate is too small, while for $e > e_1$, it is too large.

 All scalings with $m_0$, $a$, and $m_1$ have been checked and we verified BAB and ABA have no significant differences in their numerical precessions.

We can test Eq. \eqref{eq:omfin} in the context of the solar system with its current orbital parameters.  The precession associated with Mercury is $g_1 = 559 ''/\text{cent}$, while $g_5$, associated with Jupiter, is $g_5 = 426 ''/\text{cent}$.  Choosing $h = 4$ days, Eq. \eqref{eq:omfin} predicts average precession rates, which we can use as proxies for an artificial secular frequency, $\tilde{g}_i$: $\tilde{g}_1 = {-3.5 \times 10^{-7}}''   \text{/cent} $ and $\tilde{g}_5 = {-8.4 \times 10^{-5}} ''  \mathrm{/cent}$, both minuscule amounts and significantly lower than those found by \cite{Reinetal2026}.  However, it's noteworthy the impact on Jupiter is $242$ times larger than on Mercury, due to the $\epsilon^2$ effect.  
\section{General relativistic corrections and planet--planet interactions}
\label{sec:GR}
General relativity is often incorporated into solar system simulations as a position-dependent potential.  For each body,
\begin{equation}
\label{eq:Hgr}
\sub{V}{GR} = - \frac{3 G^2 m_0^2 m_1}{c^2 Q^2},
\end{equation} \par\noindent
\citep{NR86,sah92,Tamayoetal2020}, with $Q$ its heliocentric distance, and where $c$ is the speed of light.  \cite{Reinetal2026} and \cite{KaibRaymond2025} used this form.  The total Hamiltonian is now,
\begin{equation}
\label{eq:Hgrtot}
H = T_1 + T_0 + V_\odot + \sub{V}{pl} + \sub{V}{GR}.
\end{equation} \par\noindent
We can calculate the physical, not artificial, precession due to $\sub{V}{GR}$ by using Eq. \eqref{eq:preceq}.  $H$ and $\v{L}$, or $e$ and $a$, are conserved so $R$ is constant.  We use Eq. \eqref{eq:preceq} and,
\begin{subequations}
\begin{align}
\left \langle \frac{1}{Q^3} \right \rangle &= \frac{1}{a^3 (1-e^2 )^{3/2}},\\
\left \langle \frac{1}{Q^4} \right \rangle &= \frac{2 + e^2}{2 a^4 (1-e^2 )^{5/2}}.
\end{align}
\end{subequations} \par\noindent
Then, we have,
\begin{equation}
\label{eq:omgr}
\langle \dot{\varpi} \rangle= \frac{3 G^{3/2} m_0 M^{1/2}}{c^2 a^{5/2} (1-e^2)}.
\end{equation} \par\noindent
The direction is prograde, opposite the artificial contribution from earlier.  To find the size of $\sub{V}{GR}$, use the same scaling arguments as those below map \eqref{eq:mapc} to find, for Mercury,
\begin{equation}
\label{eq:GRmag}
\sub{V}{GR}/T_0 = \mathcal O \left( (v_1^2/c^2)/(m_1/m_0) \right) = 10^{-1},
\end{equation} \par\noindent
Eq. \eqref{eq:GRmag} implies we can safely place $\sub{V}{GR}$ in $\sub{B}{D}$ for solar system simulations.  

We now set up a two-body system consisting of (approximate) Mercury and Sun: $m_0 = 1$, $m_1 = 1.660 \times 10^{-7}$, and $a = 0.3871$, but $e$ is set to $0.6$ to match the second subplot of fig. 1 from \cite{Reinetal2026}.  $P = 0.2409$ and $\sub{t}{max} = 10^{5}P$.  We solve Hamiltonian \eqref{eq:Hgrtot} using BAB, comparing the numerical precession with Eq. \eqref{eq:omgr}.  $h$ is varied from $h = 0.5$ to $60$ days, again to match \cite{Reinetal2026}. 

\begin{figure}
	\includegraphics[width=80mm]{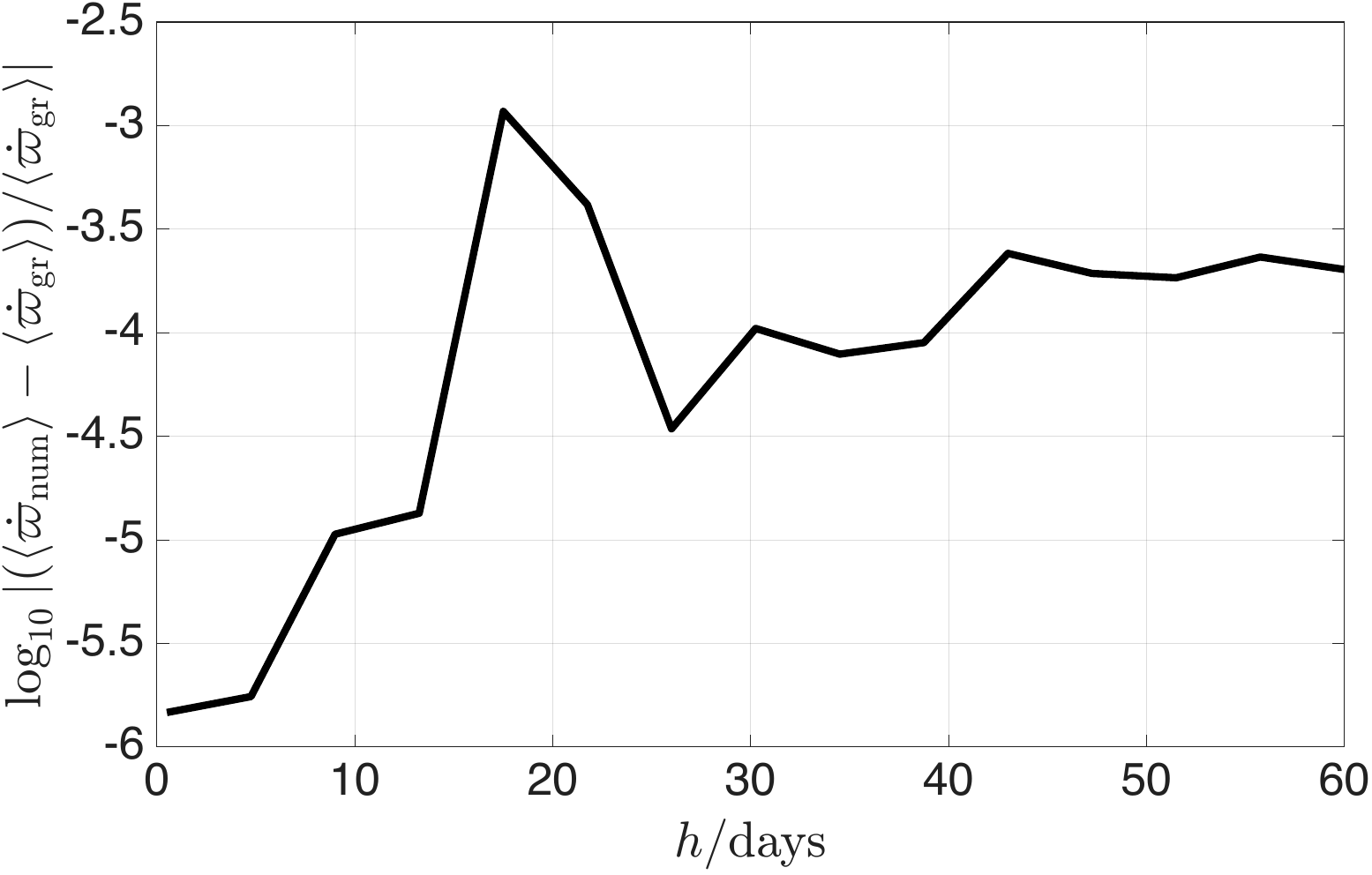}
	\caption{Error in numerical precession as compared to Eq. \eqref{eq:omgr}.  A two-body problem consisting of Mercury and Sun, with $e = 0.6$, and a GR potential, is run.  Even at large timesteps, artificial precession makes no significant impact.
	\label{fig:gr}
  	}
\end{figure}

The result is shown in Fig. \ref{fig:gr}.    
We find excellent agreement between Eq. \eqref{eq:omgr} and the numerical precession.  This contrasts with the result from \cite{Reinetal2026}, who find the precessions agree only up to $h \approx 6$ days.  The main difference between our simulations and theirs is their inclusion of $\sub{V}{pl}$, with all planets, so it is reasonable to conclude error terms, e.g., Eq. \eqref{eq:err:WHI}, that include $\sub{V}{pl}$ are responsible for their artificial precession.  $\sub{V}{pl}$, of course, is dependent on the exact planetary configuration under study.  $\sub{V}{pl}$ is responsible for most of $g_1$, so it is not unreasonable it would also be responsible for most artificial precession as well.

In fact, we're able to qualitatively reproduce fig. 1, subplot 2, from \cite{Reinetal2026}, with a simple experiment consisting of the Sun and two planets, with no GR.  All motion is in a plane, and initially the planets' arguments of pericenter are aligned and their phases are at aphelion.  Their masses and semi-major axes are those of Mercury and Jupiter, respectively, but only Jupiter's eccentricity matches its current value.  Mercury's eccentricity is set to $0.6$.  

As in Fig. \ref{fig:gr}, we run for $10^5P$, with $P$ Mercury's period.  $15$ timesteps are chosen, varying logarithmically between $0.005P$ and $0.8P$, and a simulation is run with each.  Timesteps beyond $h_0 = 1.1$ run the risk of not resolving pericenter in DHC \citep{WH92,RH99,W15,Hetal2022}, although \cite{Reinetal2025} find that larger stepsizes can still successfully estimate some dynamics.  As we know for small $h$ the artificial precession decreases with $h^2$, we choose $h = 0.005P$ as the ``analytic'' simulation.

\begin{figure}
	\includegraphics[width=80mm]{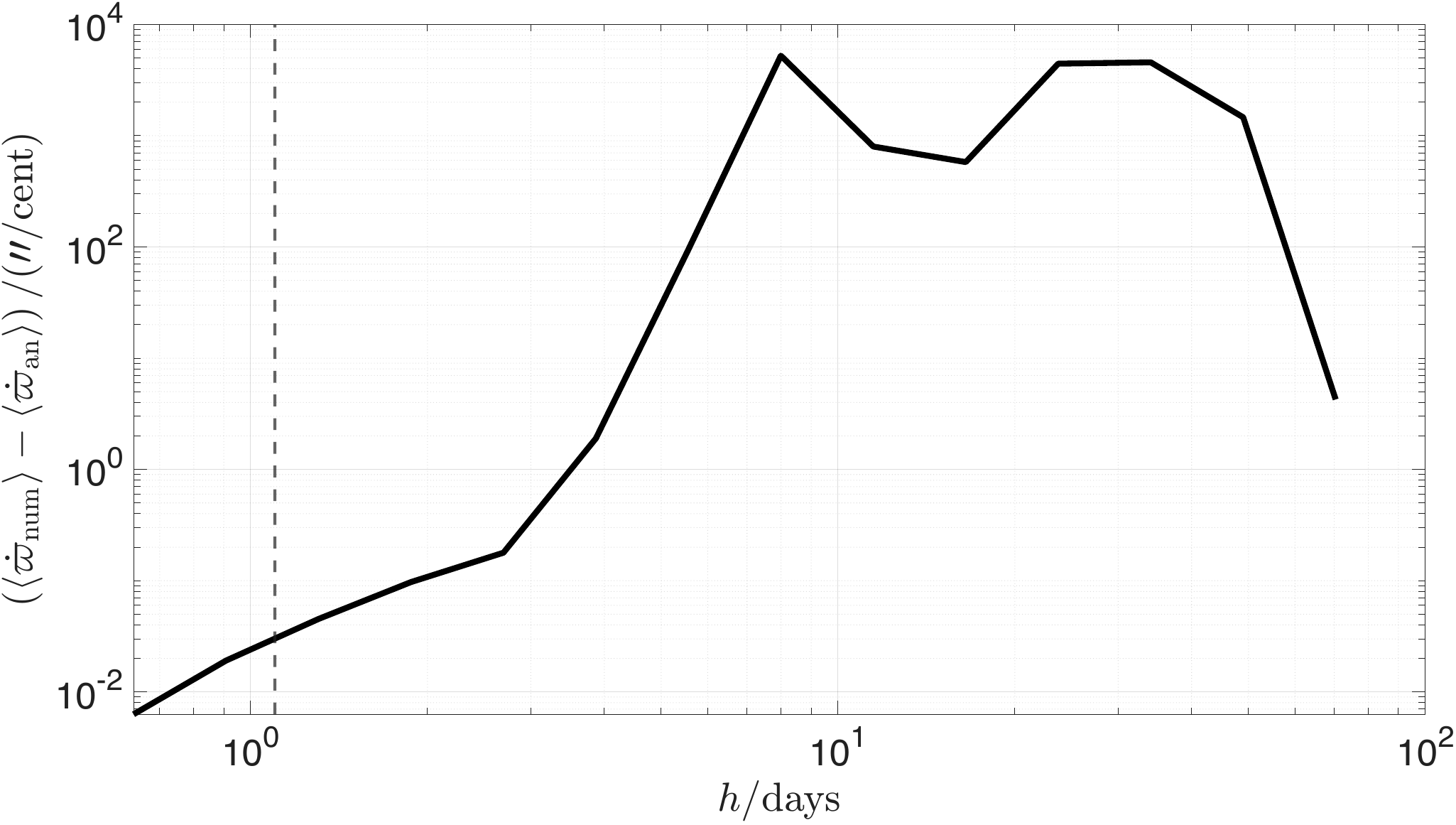}
	\caption{The artificial precession measured in a simplified two-planet Mercury--Jupiter system, as a function of timestep.  Mercury's eccentricity has been increased to $e = 0.6$, and there is no GR.  Strong prograde artificial precession for $h \ge 3.9$ days, surpassing the GR contribution if the GR interaction were included, is found.  The dashed line indicates a lower bound on the {maximum} recommended stepsize for this problem. 
	\label{fig:Vpl}
  	}
\end{figure}

In Fig. \ref{fig:Vpl} we plot the difference in average Mercury precession, $\Delta \langle \dot{\varpi} \rangle$, between the rest of simulations and the analytic run, to obtain the artificial contribution. 
As in \cite{Reinetal2026}, the artificial precession is prograde.  $\Delta \langle \dot{\varpi} \rangle \ge 95''$/cent for $h \ge 3.9$ days (except for $h = 70$ days), surpassing the GR contribution to Mercury's average precession, Eq. \eqref{eq:omgr} ($= 67''$/cent).  The slope of $\Delta \langle \dot{\varpi} \rangle$ for the $5$ data points with $h \le 2.7$ days is $2.3$, consistent within uncertainties with the expected $h^2$ scaling.  $h_0$, is again shown with a dashed line.  For $h < h_0$, artificial precession is negligible, and theory is agreeing with the numerical scaling.  In this three-body setup, Jupiter's artificial precession remains small relative to Mercury's.

\section{Discussion}
\label{sec:conc}
The Wisdom--Holman map in Democratic Heliocentric Coordinates, used extensively in studies of planetary dynamics, induces artificial precession for even a simple two-body problem.  We derived the analytic form of this precession.  Interestingly, the effects on Jupiter are far larger than on Mercury by a factor of $242$ for present-day orbital parameters.  Our two-body precession result provides the mathematical framework to understand precession in more complicated planetary systems, even when other coordinates are used.  When running simulations that include general relativity (GR), the artificial precession is negligible compared to GR precession even when using extreme stepsizes.  

Recently, \cite{Reinetal2026} found strong artificial precession for Mercury, matching the size of its GR precession, in their solar system studies.  We have reproduced the important features of their result with a simplified Mercury--Jupiter system, strongly suggesting $\sub{V}{pl}$ is responsible for our and their artificial precessions.  Analytic study of artificial precession from $\sub{V}{pl}$ depends on the specific planetary configuration.

Symplectic methods like WH are exactly Hamiltonian and are considered useful due to the surrogate Hamiltonian approximation, which states that the Hamiltonian the method solves is close to the physical Hamiltonian (e.g., Eq. \ref{eq:DHC:T}).
  Fig. \ref{fig:Vpl} shows clearly that the surrogate Hamiltonian approximation holds to only roughly $h = 2.7$ days, where the artificial precession is still small.  At larger timesteps, when artificial precession becomes large, the $h^2$ scaling of the error breaks down.  The method is no longer obeying a nearby Hamiltonian, and simulation results may become questionable.  Due to this, in this regime, one may wonder whether a symplectic method like WH is still advantageous over some non-symplectic alternatives.  A symplectic simulation like WH with large artificial precession or anomalous instability may not be being used as intended.    

Our work sheds light on the origin of artificial precession and suggests it may be an important effect on other bodies beyond Mercury.

\section{Acknowledgements}
\label{sec:ack}
We thank Richard Zeebe for suggesting the test of Fig. \ref{fig:Vpl}, Scott Tremaine for feedback, and Walter Dehnen and Ander Murua for initial encouragement.  We thank John Chambers for a thoughtful review.  We acknowledge support from the National Science and Technology Council (NSTC) in Taiwan through grant 114-2112-M-003 -020-MY3.  


\bibliography{paper}{}

@article{chan90,
	Author = {{Channell}, P.~J. and {Scovel}, C.},
	Journal = {Nonlinearity},
	Month = may,
	Pages = {231-259},
	Title = {{Symplectic integration of Hamiltonian systems}},
	Volume = 3,
	Year = 1990}

@article{GBP14,
	Archiveprefix = {arXiv},
	Author = {{Gon{\c c}alves Ferrari}, G. and {Boekholt}, T. and {Portegies Zwart}, S.~F.},
	Eprint = {1402.3325},
	Journal = {MNRAS},
	Month = may,
	Pages = {719-730},
	Primaryclass = {astro-ph.CO},
	Title = {{A Keplerian-based Hamiltonian splitting for gravitational N-body simulations}},
	Volume = 440,
	Year = 2014}

@book{SC93,
	Address = {London},
	Author = {{Sanz-Serna}, J.M. and {Calvo}, M.P.},
	Edition = {First},
	Publisher = {Chapman and Hall},
	Title = {{Numerical Hamiltonian Problems}},
	Year = 1994}

@book{hair06,
	Address = {Berlin},
	Author = {{Hairer}, E. and {Lubich}, C. and {Wanner}, G.},
	Booktitle = {Geometric Numerical Integration by E.~Hairer, C.~Lubich, and G.~Wanner.~Berlin: Springer Verlag, 2006},
	Edition = {2nd},
	Publisher = {Springer Verlag},
	Title = {{Geometrical Numerical Integration}},
	Year = 2006}

@book{leim04,
	Author = {{Leimkuhler}, B. and {Reich}, S.},
	Booktitle = {Series: Cambridge Monographs on Applied and Computational Mathematics (No.~14)},
	Publisher = {Cambridge University Press},
	Title = {{Simulating Hamiltonian Dynamics}},
	Year = 2004}

@ARTICLE{WH91,
   author = {{Wisdom}, J. and {Holman}, M.},
    title = "{Symplectic maps for the n-body problem}",
  journal = {AJ},
     year = 1991,
    month = oct,
   volume = 102,
    pages = {1528-1538},
      doi = {10.1086/115978}
}

@ARTICLE{sah92,
   author = {{Saha}, P. and {Tremaine}, S.},
    title = "{Symplectic integrators for solar system dynamics}",
  journal = {AJ},
     year = 1992,
    month = oct,
   volume = 104,
    pages = {1633-1640},
      doi = {10.1086/116347}
}

@ARTICLE{C99,
   author = {{Chambers}, J.~E.},
    title = "{A hybrid symplectic integrator that permits close encounters between massive bodies}",
  journal = {MNRAS},
     year = 1999,
    month = apr,
   volume = 304,
    pages = {793-799},
      doi = {10.1046/j.1365-8711.1999.02379.x},
   adsurl = {http://adsabs.harvard.edu/abs/1999MNRAS.304..793C}
}

@ARTICLE{DLL98,
   author = {{Duncan}, M.~J. and {Levison}, H.~F. and {Lee}, M.~H.},
    title = "{A Multiple Time Step Symplectic Algorithm for Integrating Close Encounters}",
  journal = {AJ},
     year = 1998,
    month = oct,
   volume = 116,
    pages = {2067-2077},
      doi = {10.1086/300541},
   adsurl = {http://adsabs.harvard.edu/abs/1998AJ....116.2067D}
}

@ARTICLE{LD00,
   author = {{Levison}, H.~F. and {Duncan}, M.~J.},
    title = "{Symplectically Integrating Close Encounters with the Sun}",
  journal = {AJ},
     year = 2000,
    month = oct,
   volume = 120,
    pages = {2117-2123},
      doi = {10.1086/301553},
   adsurl = {http://adsabs.harvard.edu/abs/2000AJ....120.2117L}
}

@ARTICLE{HB15,
   author = {{Hernandez}, D.~M. and {Bertschinger}, E.},
    title = "{Symplectic integration for the collisional gravitational N-body problem}",
  journal = {MNRAS},
archivePrefix = "arXiv",
   eprint = {1503.02728},
 primaryClass = "astro-ph.IM",
     year = 2015,
    month = sep,
   volume = 452,
    pages = {1934-1944},
      doi = {10.1093/mnras/stv1439},
   adsurl = {http://adsabs.harvard.edu/abs/2015MNRAS.452.1934H}
}

@ARTICLE{RH99,
   author = {{Rauch}, K.~P. and {Holman}, M.},
    title = "{Dynamical Chaos in the Wisdom-Holman Integrator: Origins and Solutions}",
  journal = {\aj},
   eprint = {astro-ph/9803340},
     year = 1999,
    month = feb,
   volume = 117,
    pages = {1087-1102},
      doi = {10.1086/300720},
   adsurl = {https://ui.adsabs.harvard.edu/abs/1999AJ....117.1087R}
}

@ARTICLE{W15,
   author = {{Wisdom}, J.},
    title = "{Resolving the Pericenter}",
  journal = {AJ},
     year = 2015,
    month = oct,
   volume = 150,
      eid = {127},
    pages = {127},
      doi = {10.1088/0004-6256/150/4/127},
   adsurl = {http://adsabs.harvard.edu/abs/2015AJ....150..127W}
}

@ARTICLE{DH17,
   author = {{Dehnen}, W. and {Hernandez}, D.~M.},
    title = "{Symplectic fourth-order maps for the collisional N -body problem}",
  journal = {\mnras},
archivePrefix = "arXiv",
   eprint = {1609.09375},
 primaryClass = "math.NA",
     year = 2017,
    month = feb,
   volume = 465,
    pages = {1201-1217},
      doi = {10.1093/mnras/stw2758},
   adsurl = {http://adsabs.harvard.edu/abs/2017MNRAS.465.1201D}
}

@ARTICLE{HD17,
   author = {{Hernandez}, D.~M. and {Dehnen}, W.},
    title = "{A study of symplectic integrators for planetary system problems: error analysis and comparisons}",
  journal = {\mnras},
archivePrefix = "arXiv",
   eprint = {1612.05329},
 primaryClass = "astro-ph.EP",
     year = 2017,
    month = jul,
   volume = 468,
    pages = {2614-2636},
      doi = {10.1093/mnras/stx547},
   adsurl = {http://adsabs.harvard.edu/abs/2017MNRAS.468.2614H}
}

@ARTICLE{Reinetal2019,
       author = {{Rein}, Hanno and {Hernandez}, David M. and {Tamayo}, Daniel and {Brown}, Garett and {Eckels}, Emily and {Holmes}, Emma and {Lau}, Michelle and {Leblanc}, R{\'e}jean and {Silburt}, Ari},
        title = "{Hybrid symplectic integrators for planetary dynamics}",
      journal = {\mnras},
         year = 2019,
        month = jun,
       volume = {485},
       number = {4},
        pages = {5490-5497},
          doi = {10.1093/mnras/stz769},
archivePrefix = {arXiv},
       eprint = {1903.04972},
 primaryClass = {astro-ph.EP},
       adsurl = {https://ui.adsabs.harvard.edu/abs/2019MNRAS.485.5490R}
}

@ARTICLE{Kinoshitaetal91,
   author = {{Kinoshita}, H. and {Yoshida}, H. and {Nakai}, H.},
    title = "{Symplectic integrators and their application to dynamical astronomy}",
  journal = {Celestial Mechanics and Dynamical Astronomy},
     year = 1991,
   volume = 50,
    pages = {59-71},
   adsurl = {https://ui.adsabs.harvard.edu/abs/1991CeMDA..50...59K}
}

@ARTICLE{Tamayoetal2020,
       author = {{Tamayo}, Daniel and {Rein}, Hanno and {Shi}, Pengshuai and {Hernandez}, David M.},
        title = "{REBOUNDx: a library for adding conservative and dissipative forces to otherwise symplectic N-body integrations}",
      journal = {\mnras},
         year = 2020,
        month = jan,
       volume = {491},
       number = {2},
        pages = {2885-2901},
          doi = {10.1093/mnras/stz2870},
archivePrefix = {arXiv},
       eprint = {1908.05634},
 primaryClass = {astro-ph.EP},
       adsurl = {https://ui.adsabs.harvard.edu/abs/2020MNRAS.491.2885T}
}

@ARTICLE{WH92,
   author = {{Wisdom}, J. and {Holman}, M.},
    title = "{Symplectic maps for the n-body problem - Stability analysis}",
  journal = {\aj},
     year = 1992,
    month = nov,
   volume = 104,
    pages = {2022-2029},
      doi = {10.1086/116378},
   adsurl = {https://ui.adsabs.harvard.edu/abs/1992AJ....104.2022W}
}

@ARTICLE{Hernandezetal2020,
       author = {{Hernandez}, David M. and {Hadden}, Sam and {Makino}, Junichiro},
        title = "{Are long-term N-body simulations reliable?}",
      journal = {\mnras},
         year = 2020,
        month = apr,
       volume = {493},
       number = {2},
        pages = {1913-1925},
          doi = {10.1093/mnras/staa388},
archivePrefix = {arXiv},
       eprint = {1910.08667},
 primaryClass = {astro-ph.EP},
       adsurl = {https://ui.adsabs.harvard.edu/abs/2020MNRAS.493.1913H}
}

@ARTICLE{Zeebe2015a,
       author = {{Zeebe}, Richard E.},
        title = "{Dynamic Stability of the Solar System: Statistically Inconclusive Results from Ensemble Integrations}",
      journal = {\apj},
         year = 2015,
        month = jan,
       volume = {798},
       number = {1},
          eid = {8},
        pages = {8},
          doi = {10.1088/0004-637X/798/1/8},
archivePrefix = {arXiv},
       eprint = {1506.07602},
 primaryClass = {astro-ph.EP},
       adsurl = {https://ui.adsabs.harvard.edu/abs/2015ApJ...798....8Z}
}

@INPROCEEDINGS{NR86,
       author = {{Nobili}, Anna and {Roxburgh}, Ian W.},
        title = "{Simulation of General Relativistic Corrections in Longterm Numerical Integrations of Planetary Orbits}",
    booktitle = {Relativity in Celestial Mechanics and Astrometry.  High Precision Dynamical Theories and Observational Verifications},
         year = 1986,
       editor = {{Kovalevsky}, Jean and {Brumberg}, V.~A.},
       volume = {114},
        month = jan,
        pages = {105},
       adsurl = {https://ui.adsabs.harvard.edu/abs/1986IAUS..114..105N}
}

@ARTICLE{Hetal2022,
       author = {{Hernandez}, David M. and {Zeebe}, Richard E. and {Hadden}, Sam},
        title = "{Stepsize errors in the N-body problem: discerning Mercury's true possible long-term orbits}",
      journal = {\mnras},
         year = 2022,
        month = mar,
       volume = {510},
       number = {3},
        pages = {4302-4307},
          doi = {10.1093/mnras/stab3664},
archivePrefix = {arXiv},
       eprint = {2111.08835},
 primaryClass = {astro-ph.EP},
       adsurl = {https://ui.adsabs.harvard.edu/abs/2022MNRAS.510.4302H}
}

@INPROCEEDINGS{Leeetal1997,
       author = {{Lee}, Man Hoi and {Duncan}, Martin J. and {Levison}, Harold F.},
        title = "{Variable Time Step Integrators for Long-Term Orbital Integrations}",
    booktitle = {Computational Astrophysics; 12th Kingston Meeting on Theoretical Astrophysics},
         year = 1997,
       editor = {{Clarke}, D.~A. and {West}, M.~J.},
       series = {Astronomical Society of the Pacific Conference Series},
       volume = {12},
        month = jan,
        pages = {32},
       adsurl = {https://ui.adsabs.harvard.edu/abs/1997ASPC..123...32L}
}

@book{blanescasas2017,
  title={A Concise Introduction to Geometric Numerical Integration},
  author={Blanes, S. and Casas, F.},
  isbn={9781482263442},
  series={Chapman \& Hall/CRC Monographs and Research Notes in Mathematics},
  year={2017},
  publisher={CRC Press}
}

@ARTICLE{HD2023,
       author = {{Hernandez}, David M. and {Dehnen}, Walter},
        title = "{Switching integrators reversibly in the astrophysical N-body problem}",
      journal = {\mnras},
         year = 2023,
        month = jul,
       volume = {522},
       number = {3},
        pages = {4639-4648},
          doi = {10.1093/mnras/stad657},
archivePrefix = {arXiv},
       eprint = {2301.06253},
 primaryClass = {astro-ph.EP},
       adsurl = {https://ui.adsabs.harvard.edu/abs/2023MNRAS.522.4639H}
}

@BOOK{Tremaine2023,
       author = {{Tremaine}, Scott},
        title = "{Dynamics of Planetary Systems}",
         year = 2023,
       adsurl = {https://ui.adsabs.harvard.edu/abs/2023dyps.book.....T}
}

@ARTICLE{Luetal2024,
       author = {{Lu}, Tiger and {Hernandez}, David M. and {Rein}, Hanno},
        title = "{TRACE: a code for time-reversible astrophysical close encounters}",
      journal = {\mnras},
         year = 2024,
        month = sep,
       volume = {533},
       number = {3},
        pages = {3708-3723},
          doi = {10.1093/mnras/stae1982},
archivePrefix = {arXiv},
       eprint = {2405.03800},
 primaryClass = {astro-ph.EP},
       adsurl = {https://ui.adsabs.harvard.edu/abs/2024MNRAS.533.3708L}
}

@ARTICLE{Reinetal2026,
       author = {{Rein}, Hanno and {Dey}, Kavi and {Tamayo}, Daniel},
        title = "{Democratic heliocentric coordinates underestimate the rate of instabilities in long-term integrations of the Solar System}",
      journal = {arXiv e-prints},
         year = 2026,
        month = jan,
          eid = {arXiv:2601.08019},
        pages = {arXiv:2601.08019},
          doi = {10.48550/arXiv.2601.08019},
archivePrefix = {arXiv},
       eprint = {2601.08019},
 primaryClass = {astro-ph.EP},
       adsurl = {https://ui.adsabs.harvard.edu/abs/2026arXiv260108019R}
}

@ARTICLE{Javaherietal2023,
       author = {{Javaheri}, Pejvak and {Rein}, Hanno and {Tamayo}, Dan},
        title = "{WHFast512: A symplectic N-body integrator for planetary systems optimized with AVX512 instructions}",
      journal = {The Open Journal of Astrophysics},
         year = 2023,
        month = jul,
       volume = {6},
          eid = {29},
        pages = {29},
          doi = {10.21105/astro.2307.05683},
archivePrefix = {arXiv},
       eprint = {2307.05683},
 primaryClass = {astro-ph.EP},
       adsurl = {https://ui.adsabs.harvard.edu/abs/2023OJAp....6E..29J}
}

@ARTICLE{Grimmetal2014,
       author = {{Grimm}, Simon L. and {Stadel}, Joachim G.},
        title = "{The GENGA Code: Gravitational Encounters in N-body Simulations with GPU Acceleration}",
      journal = {\apj},
         year = 2014,
        month = nov,
       volume = {796},
       number = {1},
          eid = {23},
        pages = {23},
          doi = {10.1088/0004-637X/796/1/23},
archivePrefix = {arXiv},
       eprint = {1404.2324},
 primaryClass = {astro-ph.EP},
       adsurl = {https://ui.adsabs.harvard.edu/abs/2014ApJ...796...23G}
}

@ARTICLE{KaibRaymond2025,
       author = {{Kaib}, Nathan A. and {Raymond}, Sean N.},
        title = "{The influence of passing field stars on the solar system's dynamical future}",
      journal = {Icarus},
         year = 2025,
        month = oct,
       volume = {439},
          eid = {116632},
        pages = {116632},
          doi = {10.1016/j.icarus.2025.116632},
archivePrefix = {arXiv},
       eprint = {2505.04737},
 primaryClass = {astro-ph.EP},
       adsurl = {https://ui.adsabs.harvard.edu/abs/2025Icar..43916632K}
}

@ARTICLE{Reinetal2025,
       author = {{Rein}, Hanno and {Brown}, Garett and {Kanda}, Mei},
        title = "{On the statistical convergence of N-body simulations of the Solar System}",
      journal = {The Open Journal of Astrophysics},
         year = 2025,
        month = dec,
       volume = {8},
        pages = {54745},
          doi = {10.33232/001c.154745},
archivePrefix = {arXiv},
       eprint = {2507.04987},
 primaryClass = {astro-ph.EP},
       adsurl = {https://ui.adsabs.harvard.edu/abs/2025OJAp....854745R}
}

@ARTICLE{Campbell1996,
   author = {{Campbell}, J.E.},
    title = "{On a Law of Combination of Operators bearing on the Theory of Continuous Transformation Groups}",
     year = {1896},
   volume = {s1-28},
    pages = {381-390},
      doi = {10.1112/plms/s1-28.1.381},
      URL = {http://plms.oxfordjournals.org/content/s1-28/1/381.short}, 
   eprint = {http://plms.oxfordjournals.org/content/s1-28/1/381.full.pdf+html}, 
  journal = {Proc.\ London Math.\ Soc.},
}

@ARTICLE{Campbell1997,
   author = {{Campbell}, J.E.},
    title = "{On a Law of Combination of Operators (Second Paper)}",
     year = {1897},
   volume = {s1-29},
    pages = {14-32},
      doi = {10.1112/plms/s1-29.1.14},
      URL = {http://plms.oxfordjournals.org/content/s1-29/1/14.short}, 
   eprint = {http://plms.oxfordjournals.org/content/s1-29/1/14.full.pdf+html}, 
  journal = {Proc.\ London Math.\ Soc.},
}

@article{Baker1905,
   author = {{Baker}, H. F.}, 
    title = {Alternants and Continuous Groups},
   volume = {s2-3}, 
    pages = {24-47}, 
     year = {1905}, 
      doi = {10.1112/plms/s2-3.1.24}, 
      URL = {http://plms.oxfordjournals.org/content/s2-3/1/24.short}, 
   eprint = {http://plms.oxfordjournals.org/content/s2-3/1/24.full.pdf+html}, 
  journal = {Proc.\ London Math.\ Soc.} 
}

@article{Baker1902b,
   author = {{Baker}, H. F.}, 
    title = "{On the Integration of Linear Differential Equations}",
   volume = {s1-35}, 
    pages = {333-378}, 
     year = {1902}, 
      doi = {10.1112/plms/s1-35.1.333}, 
      URL = {http://plms.oxfordjournals.org/content/s1-35/1/333.short}, 
   eprint = {http://plms.oxfordjournals.org/content/s1-35/1/333.full.pdf+html}, 
  journal = {Proc.\ London Math.\ Soc.} 
}

@article{Hausdorff1906,
   author = {{Hausdorff}, F.}, 
    title = "{Die symbolische Exponentialformel in der Gruppentheorie}",
   volume = {58}, 
    pages = {19-48}, 
     year = {1906}, 
  journal = {Ber.\ Verh.\ S\"a{}chs.\ Adak.\ Wiss.\} 
}

@article{Dynkin1947,
   author = {{Dynkin}, E.B.},
  journal = {Doklady Akademii Nauk SSSR},
   volume = {57},
    pages = {323-326},
     year = 1947,
}

@ARTICLE{Abbotetal2024,
       author = {{Abbot}, Dorian S. and {Webber}, Robert J. and {Hernandez}, David M. and {Hadden}, Sam and {Weare}, Jonathan},
        title = "{Mercury's Chaotic Secular Evolution as a Subdiffusive Process}",
      journal = {\apj},
         year = 2024,
        month = jun,
       volume = {967},
       number = {2},
          eid = {121},
        pages = {121},
          doi = {10.3847/1538-4357/ad3e5f},
archivePrefix = {arXiv},
       eprint = {2306.11870},
 primaryClass = {astro-ph.EP},
       adsurl = {https://ui.adsabs.harvard.edu/abs/2024ApJ...967..121A}
}

@article{Fasso2005,
  title={Superintegrable Hamiltonian Systems: Geometry and Perturbations},
  author={Fass{\`o}, Francesco},
  journal={Acta Applicandae Mathematica},
  volume={87},
  number={1-3},
  pages={101--111},
  year={2005},
  publisher={Springer},
  doi={10.1007/s10440-005-1139-8}
}

@INPROCEEDINGS{RauchHamilton2002,
       author = {{Rauch}, K.~P. and {Hamilton}, D.~P.},
        title = "{The HNBody Package for Symplectic Integration of Nearly-Keplerian Systems}",
    booktitle = {AAS/Division of Dynamical Astronomy Meeting \#33},
         year = 2002,
       series = {AAS/Division of Dynamical Astronomy Meeting},
       volume = {33},
        month = sep,
          eid = {08.02},
        pages = {08.02},
       adsurl = {https://ui.adsabs.harvard.edu/abs/2002DDA....33.0802R}
}
\bibliographystyle{aasjournalv7}

\end{document}